# Suppression of Ferromagnetic Double Exchange by Vibronic Phase Segregation

F. Rivadulla[1], M. Otero-Leal[2], A. Espinosa[3], A. de Andrés[3], C. Ramos[4], J. Rivas[2], J. B. Goodenough[5]

[1]Physical-Chemistry and [2]Applied Physics Departments, University of Santiago de Compostela, 15782 Santiago de Compostela, Spain.
[3]Instituto de Ciencia de Materiales de Madrid, Consejo Superior de Investigaciones Científicas, Cantoblanco, E-28049 Madrid, Spain
[4]Centro Atómico Bariloche and Instituto Balseiro, 8400 San Carlos de Bariloche, Rio Negro, Argentina
[5]Texas Materials Institute, ETC 9.102, The University of Texas at Austin, 1 University Station, C2201, Austin, Texas 78712, USA

From Raman spectroscopy, magnetization, and thermal-expansion on the system $La_{2/3}(Ca_{1-x}Sr_x)_{1/3}MnO_3$, we have been able to provide a quantitative basis for the heterogeneous electronic model for manganites exhibiting colossal magnetoresistance (CMR). We construct a mean-field model that accounts quantitatively for the measured deviation of $T_C(x)$ from the $T_C$ predicted by de Gennes double exchange in the adiabatic approximation, and predicts the occurrence of a first order transition for a strong coupling regime, in accordance with the experiments. The existence of a temperature interval $T_C<T<T^*$ where CMR may be found is discussed, in connection with the occurrence of an idealized Griffiths phase.

The $La_{1-y}Sr_yMnO_3$ perovskite system with y>0.17 has been shown[1] to be a de Gennes[2] double-exchange ferromagnet; it contains itinerant electrons in a narrow $\sigma^*$ band of e-orbital parentage coexisting with a localized $t^3$ configuration at the high-spin $Mn^{4+}/Mn^{3+}$ mixed-valent $MnO_3$ array. The adiabatic approximation, which assumes the electrons and the ions make a separate contribution to the total energy of the $MnO_3$ array, is applicable to the itinerant electrons of $La_{0.7}Sr_{0.3}MnO_3$. On the other hand, early experiments[3] on the system $(La_{1-x}Pr_x)_{0.7}Ca_{0.3}MnO_3$ showed a remarkable decrease of $T_C$ and a colossal magnetoresistance (CMR) effect above $T_C$ that increased with decreasing x until a transition to an orbitally ordered, antiferromagnetic-insulator phase. This striking evolution with x of magnetic and transport properties was interpreted[4] to result from the stabilization of phase fluctuations of itinerant and localized electronic regions due to lattice instabilities associated with a first-order transition at the crossover from polaronic to itinerant electron behavior. Although evidences for the phase fluctuations and their responsibility for the CMR phenomenon have been well-documented, a quantitative description of the evolution of $T_C$ with x has been lacking. In order to fill this gap, we have undertaken a systematic study of the ferromagnetic system $La_{2/3}(Ca_{1-x}Sr_x)_{1/3}MnO_3$ which shows a similar CMR phenomenon and variation of $T_C$ with x. The study allows us to account quantitatively for the evolution of $T_C$ with x using experimentally determined parameters. We demonstrate that dynamic phase segregation into hole-rich itinerant-electron regions and hole-poor localized-electron regions, results from a breakdown of the adiabatic approximation for the itinerant electrons and suppresses the conventional double-exchange coupling. Our model provides an original scenario to understand the first order nature of the magnetic phase transition in a situation in which $T_C$ is suppressed, which could be of general applicability to other systems.

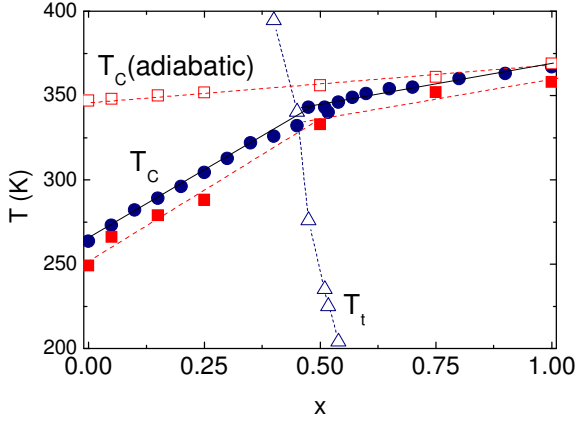

Figure 1: (solid circles): evolution of $T_C$ with x across the series $La_{2/3}(Ca_{1-x}Sr_x)_{1/3}MnO_3$ determined experimentally. (open triangles): $Pnma \leftrightarrow R\bar{3}c$ structural transition temperature determined experimentally. (open squares): evolution of $T_C$ with x calculated from a free-electron DE model (equation (1)), as explained in the text. (solid squares): calculated $T_C$ with the mean-field model coupled to the lattice proposed in this work as explained in the text.

Fig. 1 shows the experimental magnetic and structural phase diagram for $La_{2/3}(Ca_{1-x}Sr_x)_{1/3}MnO_3$. The progressive introduction of Ca in this system increases the bending of the (180°-ϕ) Mn-O-Mn bond angle,[5] which induces a phase transition from rhombohedral ($R\bar{3}c$) to orthorhombic (Pbnm) symmetry at a room-temperature critical concentration $x_t \approx 0.45$. In the adiabatic approximation, the ferromagnetic Curie temperature $T_C$ is proportional to the double-exchange energy parameter

$$J \sim c(1-c)zb\cos\phi\cos(\theta_{ij}/2) \qquad (1)$$

where c and (1−c) are the fraction of $Mn^{3+}$ and $Mn^{4+}$ states at the z nearest neighbors to which an electron can hop from $Mn^{3+}$ to $Mn^{4+}$, b is the conventional spin-independent energy-transfer integral, and $\theta_{ij}$ is the angle between the spins on the two ions where electron transfer occurs.[6] The open squares in Fig. 1 ($T_C$(adiabatic)) are an estimation of the $T_C$ vs. $\cos\phi$ to be expected for a homogenous system of itinerant $\sigma^*$ electrons as extrapolated from the value of $T_C$ for $La_{2/3}Sr_{1/3}MnO_3$. The deviation of the experimental $dT_C/dx$ from this extrapolation is less than 4% in the rhombohedral phase; but in the orthorhombic phase the deviation is much more pronounced. For the end member $La_{2/3}Ca_{2/3}MnO_3$ (x=0), the discrepancy in $T_C$ is nearly 100 K. Inclusion of the effect of a long-range stress field[7] or of the variance in the A-site cation radii[8] cannot explain this strong suppression of $T_C$ or its variation with x. The more rapid decrease of $T_C$ with x from the extrapolated value in the orthorhombic phase is suggestive that the adiabatic approximation for the narrow $\sigma^*$ band breaks down more rapidly in this phase than in the rhombohedral phase. This can be qualitatively understood: narrowing of the adiabatic $\sigma^*$ bandwidth $W_0 \sim b\cos\phi$ by increasing the bending $\phi$ of the (180°-$\phi$) Mn-O-Mn angles increases the time $\tau_h \sim h/W_0$ for an electron to tunnel from a $Mn^{3+}$ to a $Mn^{4+}$ ion, and the adiabatic approximation is only valid for $\tau_h < \omega_o^{-1}$, where $\omega_o^{-1}$ is the period of the optical-mode vibration of the $MnO_3$ array that would localize the electron as a small polaron (or as a two-manganese Zener polaron). Where a $\tau_h \approx \omega_o^{-1}$ is approached, $\tau_h$ increases with the depth of the polaronic trap state, and this trap state is deepened by a Jahn-Teller deformation (orbital ordering) of the site. The orthorhombic crystal symmetry allows a Jahn-Teller deformation whereas rhombohedral symmetry does not support lifting of the e-orbital degeneracy. Zhao et al.[9] reported a large oxygen isotope effect on $T_C$ of orthorhombic manganites where long-range Janh-Teller distortions are present. Increasing the mass of the oxygen isotope decreases the frequencies of the Mn-O-Mn modes and $T_C$ decreases. If on cooling through $T_C$ the transition was from a global polaronic phase ($\tau_h > \omega_o^{-1}$) to a global itinerant electronic phase ($\tau_h < \omega_o^{-1}$), then the exchange of $^{18}O$ for $^{16}O$ should favor the itinerant electron phase. The decrease of $T_C$ by $^{18}O/^{16}O$ exchange was interpreted[4] as an evidence of the two-phase character of the electronic system. The dramatic changes in $T_C$ by Mn-O-Mn angle variation should be then caused by the dependence of volume fraction of the polaronic/itinerant phases on the $\omega_o(\phi)$. In order to probe whether such a relation exists, we measured with Raman spectroscopy the evolution with x of the frequency of the oxygen vibrations perpendicular to a Mn-O-Mn bond.[10] Fig. 2 shows a linear decrease in this frequency, *i.e.* an increase in $\omega_o^{-1}$, with increasing Sr.

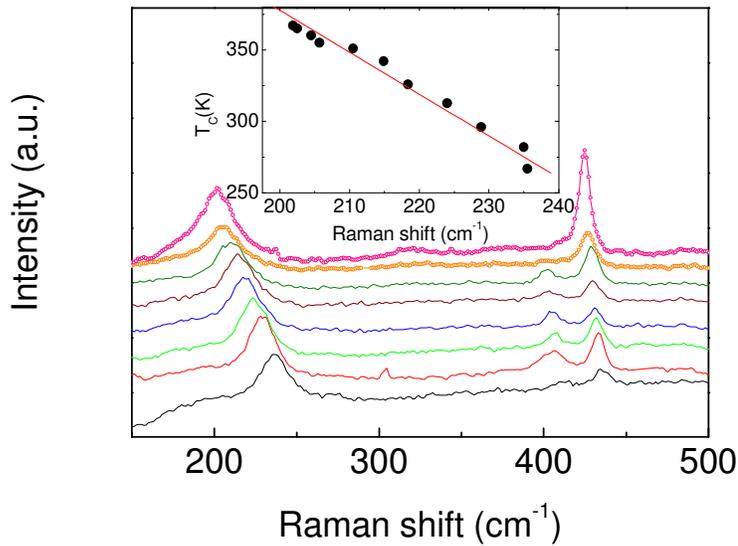

Figure 2: Raman spectra for $La_{2/3}(Ca_{1-x}Sr_x)_{1/3}MnO_3$ at 100 K. From bottom to top x=0, 0.20, 0.30, 0.40, 0.50, 0.60, 0.65, 1. The last two, plotted with open circles, belong to the $R\bar{3}c$ group at this temperature (note that at 100 K the orthorhombic-to-rhombohedral transition occurs for x≈0.6, see Fig. 1). Inset: Dependence of the $T_C$ with the frequency of the Mn-O-Mn tilting mode.

The similarity of $dT_C/dx$ and $d\omega_o^{-1}/dx$ is remarkable, providing a solid basis for the heterogeneous model. A system which is at a crossover from polaronic to itinerant electronic behavior is intrinsically unstable relative to segregation into hole-rich, itinerant-electron regions with $\tau_h < \omega_o^{-1}$ and hole-poor polaronic regions with $\tau_h > \omega_o^{-1}$. Since the phase segregation is dynamic, the measured $\omega_o$ reflects a weighted average of the two frequencies for the two volume fractions; the Raman peak is not split into two peaks as would be expected for classic, static phase segregation. However, the peak shifts to higher frequencies as the volume fraction of the polaronic phase increases.[11] With this interpretation of the Raman data of Fig. 2, we conclude that the deviation of $T_C$ from the predicted by the adiabatic approximation at low x reflects a continuous growth of the volume fraction of the polaronic phase as $\phi$ increases. In the polaronic phase, $\tau_h$ becomes too long relative to the spin reorientation time ($\tau_h > \tau_r$) for the double-exchange mechanism to be operational. In the system $La_{1-x}Sr_xMnO_3$, $T_C$ was reported to increase rapidly with pressure for the interval $0.1<x<0.2$,[12] where a short-range dynamic ordering at Mn(III) ions introduces a modified version of Zener's[13] ferromagnetism. On crossing to the high-pressure itinerant side through a first-order transition, de Gennes model is restored and $dT_C/dP$ decreases. These results are in perfect accordance with our experimental observations.

Fig. 3 shows the temperature dependence of ZFC and FC curves for x = 0.5; this sample undergoes an orthorhombic to rhombohedral transition at $T_t$ = 225 K < $T_C$. Both the splitting of the ZFC and FC curves in the orthorhombic phase as well as its lower magnetization relative to that of the rhombohedral region, are clear indications of an inhomogeneous magnetic system, even well below $T_C$. This, along with the departure of $T_C$ from the adiabatic prediction and the Raman experiment, strongly support a dynamic coexistence of two phases in the orthorhombic phase, [14] the itinerant-electron volume fraction increasing in an applied magnetic field. On the other hand the behavior in the rhombohedral phase is characteristic of a homogeneous itinerant-electron phase.

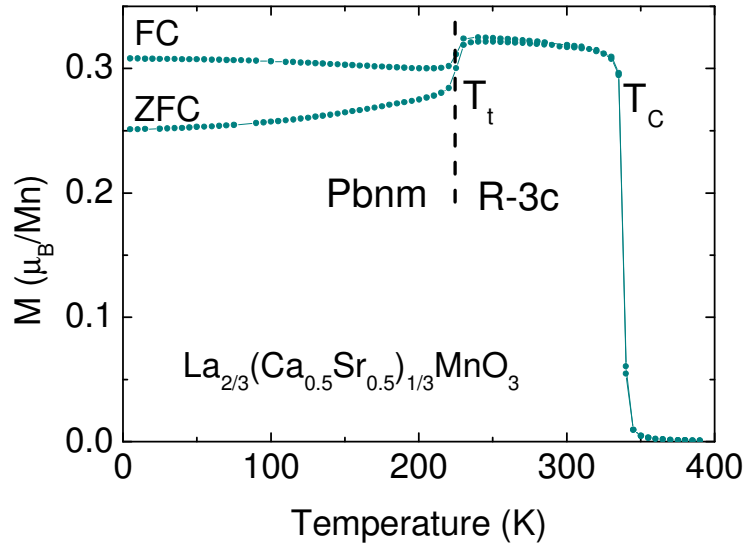

Figure 3: Temperature dependence of the ZFC-FC magnetization curves of x=0.5 measured at 100 Oe.

In order to translate the implications of an heterogeneous electronic model to a quantitative calculation of $T_C$ vs. x, we constructed a mean-field model starting from the equation of state of a ferromagnet, in which $\sigma(T) = M(T)/M(0)$, has the form

$$\sigma = B(y); \qquad y = 2zJS^2\sigma/kT \qquad (2)$$

and determine how the double-exchange energy parameter J entering the Brillouin function B(y) varies with x, *i.e.* J = J(x). The value of J will depend on the probability that a $\tau_h < \omega_o^{-1}$ to a neighboring $Mn^{4+}$ ion can occur, *i.e.* to the volume fraction of itinerant electrons. If n is the volume fraction of itinerant electrons, then J in equation (2) should be replaced by

$$J = nJ^0, \text{ where } n = (1 - n_{JT}) \qquad (3)$$

$J^0$ would be the exchange parameter estimated from the adiabatic model and $n_{JT}$ is the volume fraction of electrons in the polaronic phase.

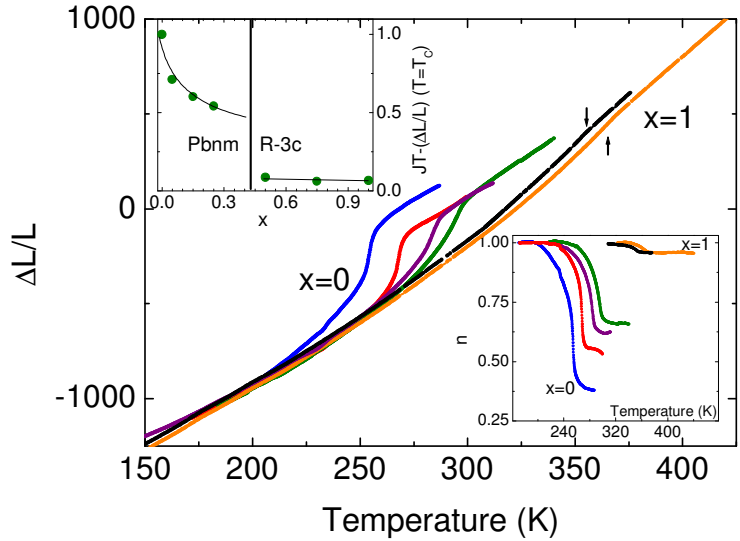

Figure 4: Linear thermal expansion of some representative composition of the series $La_{2/3}(Ca_{1-x}Sr_x)_{1/3}MnO_3$ (x=0, 0.05, 0.15, 0.25, 0.75, and 1, from left to right). The arrows mark the $T_C$ in the rombohedral samples, where the anomalous contribution due to JT distortion is suppressed. Upper inset: Relative contribution of the anomalous thermal expansion at $T_C$. The values where normalized with respect to the contribution at the x=0 sample. Lower inset: Temperature dependence of the number of delocalised electrons, n, extracted from the thermal expansion experiments as explained in the text.

Since the volume of the polaronic phase is greater than that of the itinerant-electron phase[4], a value of $n_{JT}$ can be derived from the deviation of the thermal expansion from the Grüneisen law. Fig. 4 shows the lattice thermal expansion for different representative samples of $La_{2/3}(Ca_{1-x}Sr_x)_{1/3}MnO_3$. For x=0 there is an abrupt departure from the Grüneisen contribution and an anomalous volume expansion at $T_C$ as has been previously observed by other authors.[15] With increasing the Sr content in the orthorhombic phase, this anomaly decreases and it drops abruptly on entering the rhombohedral phase where it is nearly fully suppressed (see Fig.4, upper inset). From numerous studies of the CMR phenomenon it has been established that well-below $T_C$ nearly the entire volume is itinerant, and above $T_C$ the CMR compound becomes biphasic with a sharp change in the relative volume fractions on crossing $T_C$. For this reason, the values of $n_{JT}$ derived from $\Delta L/L$ for x=0 were normalized to vary between 0 at low temperatures (where almost all the electrons are free) and 2/3 above $T_C$, were the all the electrons form polarons (there are only 2/3 of the Mn sites occupied by $Mn^{3+}$). The experimental results for $n=1-n_{JT}$ are shown in the lower inset of Fig. 4. The results for all the samples were normalized with respect to x=0, in order to establish a comparison on the evolution of the number of localized sites across the series. The continuous departure of n from 1 as $T_C$ is approached from below reflects the progressive appearance of JT-distorted sites, as it was demonstrated for similar compositions by more local probes.[14] The progressive reduction of the volume fraction of the polaronic phase as x increases is in good accord with the reduction in the frequency of the tilting mode observed by Raman. The normalized magnetization $\sigma(T)$ for different values of x according to eqs. (2) and (3) are shown in Fig. 5. At $La_{2/3}Ca_{1/3}MnO_3$, the continuous renormalization of the double-exchange interaction according to equation (3) reduces $T_C$ by up to 30% from the value of $T_C$ for the adiabatic double-exchange mechanism ($J=J^0$). The values of $T_C(x)$ estimated from eq. (2) and the measured $n(T,x)=(1-n_{JT})$, are shown in Fig. 1 (as the solid squares) for the series $La_{2/3}(Ca_{1-x}Sr_x)_{1/3}MnO_3$. An excellent quantitative agreement with the experimental $T_C(x)$ is obtained.

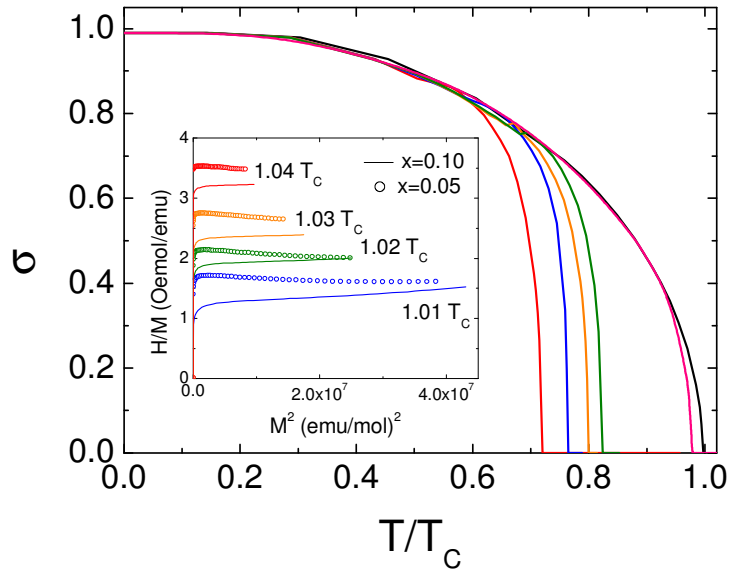

Figure 5: Evolution of the reduced magnetization according to eq. (2) and (3). Inset: Experimental isotherms probing the first-to-sencond order change in the character of the magnetic phase transition at $T_C$, on going from x=0.05 to x=0.10.

In addition, where the change in $n_{JT}(T)$, *i.e.* the fraction of sites excluded from the double-exchange coupling, is very abrupt on approaching $T_C$ from below, the model predicts the occurrence of a first-order magnetic phase transition at $T_C$. The abrupt renormalization of the exchange parameter on the approach to $T_C$ makes it impossible for the Brillouin function to evolve continuously to zero, and it collapses in a first-order phase change at the critical point. Therefore, as the volume fraction of the polaronic phase at $T_C$ increases, we can anticipate the magnetic-ordering transition at $T_C$ will change from second-order to first-order. From M/H vs. $M^2$ isotherms[16,17] we have determined experimentally that such a change does indeed occur, and takes place in the interval 0.05<x<0.10 (see inset of Fig. 5). It should be noted that the metal-to-semiconducting transition at $T_C$ stands up to x=0.5, where it goes metal-to-metal.[18] This shows that the change in the order of the phase transition is not controlled by the nature of the electronic transition, but by the two-phase character of the system and its relative volume change at $T_C$.

Previous attempts to account for first-order ferromagnetic-to-paramagnetic transitions normally relied on a volume dependent exchange interaction in a deformable lattice.[19] This ends up with a first-order phase change at a higher temperature than the corresponding second-order one in the non-deformable lattice, and so they are inadequate to explain the behavior of manganites, where $T_C$ is suppressed. The continuous change in the effective temperature of the spin-lattice of our model mimics the effect of a temperature dependent Weiss field coefficient, which also produces a first-order phase transtition.[20] Also Huberman and Streifer[21] proposed a similar model to account for the first-order character of the magnetic transition at $T_C$ in MnBi.

In conclusion, we have provided a quantitative basis for the heterogeneous model [4] responsible for the CMR phenomenon. The model can account quantitatively for the evolution of $T_C(x)$ and the change from a first to a second-order magnetic transition at $T_C$. In addition, in the temperature interval $T_C<T<T_C$(adiabatic), the model predicts pockets of short-range ferromagnetic order to exist in a non-percolative volume fraction of the itinerant-electron phase; this volume fraction grows in an applied magnetic field to beyond percolation to give the CMR phenomenon. This shares some similarities with the idealized model proposed by Griffiths[22,23] in which an exchange parameter varies with the probable number of nearest neighbors participating in the exchange interaction. However Griffiths did not anticipated how this probability might change as a function of temperature, as it does in manganites. Finally, it should be recognized that Dagotto[24] has independently supported a similar heterogeneous model developed from computational studies, and that there is a strong similarity between $T_C$(adiabatic) and the temperature scale represented by $T^*$ that he proposed. The analogy between $T^*$ and the Griffiths temperature was also raised by these authors,[25] which stressed the relevance of having two competing orders with similar energies to observe the intrinsically inhomogeneous regime.

**Acknowledgements**

We are thankful to E. Dagotto, G. Martinez, J. Mira, J.-S. Zhou for valuable comments and reading of the manuscript, and C. Hoppe and I. Pardiñas for assistance. We also thank support from MEC of Spain (MAT2002-11850-E; MAT 2004-05130-C01). FR acknowledges MCYT of Spain for support under program Ramón y Cajal.

[1] J.-S. Zhou, J. B. Goodenough, A. Asamitsu, Y. Tokura, Phys. Rev. Lett. **79**, 3234 (1997).
[2] P. G. de Gennes, Phys. Rev. **118**, 141 (1960).
[3] H. Y. Hwang, S-W. Cheong, P. G. Radaelli, M. Marezio, B. Batlogg, Phys. Rev. Lett. **75**, 914 (1995).
[4] J. B. Goodenough and J.-S. Zhou, "Localizad to Itinerant Electronic Transition in Perovskite Oxides", Structure and Bonding vol. 98. Ed. by J. B. Goodenough, Springer 2001. Chapter 2.
[5] P.G. Radaelli, G. Iannone, M. Marezio, H. Y. Hwang, S.-W. Cheong, J. D. Jorgensen, D. N. Argyriou, Rev. Lett. **56**, 8265 (1997).
[6] J. B. Goodenough, in "The Physics of Manganites", Ed. by Kaplan and Mahanti, Kluwer Academic/Plenum Pub. 1999.
[7] T. Egami, D. Louca, Phys. Rev. B **65**, 94422 (2001).
[8] J. P. Attfield, A. L. Kharlanov, J. A. McAllister, Nature **394**, 157 (1998).
[9] G.-M. Zhao, K. Konder, H. Keller, K. A. Müller, Nature **381**, 676 (1996).
[10] L. Martín-Carrón, A. de Andrés, M. J. Martínez-Lope, M. T. Casais, J. A. Alonso, Phys. Rev. B **66**, 174303 (2002).
[11] We have checked that there is a small but visible contribution of the polaronic phase at $T>T_C$ to the bending mode at $\approx 230$ cm$^{-1}$. This is in accordance with the results presented in ref. 10.

[12] J.-S. Zhou, J. B. Goodenough, Phys. Rev. B **62**, 3834 (2000).

[13] C. Zener, Phys. Rev. **82**, 403 (1951).

[14] S. J. L. Billinge, Th. Proffen, V. Petkov, J. L. Sarrao, S. Kycia, Phys. Rev. B **62**, 1203-1211 (2000).

[15] J. M. De Teresa *et al*., Nature **386**, 256 (1997).

[16] H. E. Stanley, "Introduction to Phase Transitions and Critical Phenomena", Oxford University Press, 1971.

[17] J. Mira, J. Rivas, F. Rivadulla, C. Vázquez-Vázquez, M. A. López-Quintela, Phys. Rev. B **60**, 2998 (1999).

[18] Y. Tomioka, A. Asamitsu, Y. Tokura, Phys. Rev. B **63**, 024421 (2001).

[19] C. P. Bean, D. S. Rodbell, Phys. Rev. **126**, 104 (1962).

[20] J. S. Smart, Phys. Rev. **90**, 55 (1953).

[21] B. A. Huberman, W. Streifer, Phys. Rev. B **12**, 2741 (1975).

[22] R. B. Griffiths, Phys. Rev. Lett. **23**, 17 (1969).

[23] M. B. Salamon, P. Lin, and S. H. Chun, Phys. Rev. Lett. **88**, 197203 (2002).

[24] E. Dagotto "Nanoscale Phase Separation and Colossal Magnetoresistance" Springer 2003.

[25] J. Burgy, M. Mayr, V. Martín-Mayor, A. Moreo, E. Dagotto, Phys. Rev. Lett. **87**, 277202 (2001).